\documentclass{article}

\usepackage{times}

\usepackage{graphicx}
\usepackage{subfigure}

\usepackage{natbib}

\usepackage{algorithm}
\usepackage{algorithmic}

\usepackage{hyperref}

\usepackage[accepted]{icml2014}

\usepackage{amsfonts}
\newcommand\given[1][]{\:#1\vert\:}

\icmltitlerunning{Parametric Inference using Persistence Diagrams}

\begin{document}

\twocolumn[
\icmltitle{Parametric Inference using Persistence Diagrams: \\
A Case Study in Population Genetics}

\icmlauthor{Kevin Emmett}{kje2109@columbia.edu}
\icmladdress{Columbia University,
            New York, NY.}
\icmlauthor{Daniel Rosenbloom}{dsr2131@columbia.edu}
\icmladdress{Columbia University,
            New York, NY.}
\icmlauthor{Pablo Camara}{pablo.g.camara@gmail.com}
\icmladdress{University of Barcelona,
			Barcelona, Spain.}
\icmlauthor{Raul Rabadan}{rr2579@c2b2.columbia.edu}
\icmladdress{Columbia University,
            New York, NY.}
% \icmladdress{Department of Systems Biology,
%             Columbia University,
%             New York, NY.}

% You may provide any keywords that you
% find helpful for describing your paper; these are used to populate
% the "keywords" metadata in the PDF but will not be shown in the document
\icmlkeywords{persistent homology, coalescent, population genetics}

\vskip 0.3in
]

\begin{abstract}

Persistent homology computes topological invariants from point cloud data.
Recent work has focused on developing statistical methods for data analysis in this framework.
We show that, in certain models, parametric inference can be performed using statistics defined on the computed invariants.
We develop this idea with a model from population genetics, the coalescent with recombination.
We apply our model to an influenza dataset, identifying two scales of topological structure which have a distinct biological interpretation.

\end{abstract}

\section{Introduction}
\label{sec:introduction}

Computational topology is emerging as a new approach to data analysis, driven by efficient algorithms for computing topological structure in data.
Perhaps the most mature tool is persistent homology, which summarizes multiscale topological information in a two-dimensional persistence diagram (see Figure \ref{fig:persistence_diagram} and Section \ref{sec:background}).
Recent work has concentrated on developing the statistical foundations for data analysis using the persistent homology framework \cite{Balakrishnan:2013tg,Blumberg:2012ub,Chazal:2014vl}.
The focus of this work has been estimating the topology of an object from a finite, noisy sample.
Doing so requires statistical methods to distinguish topological signal from noise.

Here we consider a different scenario.
Many simple stochastic models generate complex data that cannot be readily visualized as a manifold or summarized by a small number of topological features.
These models will generate persistence diagrams whose complexity increases with the number of sampled points.
Nevertheless, the collection of measured topological features may exhibit additional structure, providing useful information about the underlying data generating process.
While the persistence diagram is itself a summary of the topological information contained in a sampled point cloud, to perform inference further summarization may be appropriate, e.g. by considering distributions of properties defined on the diagram.
In other words, we are less interested in learning the topology of a particular sample, but rather in understanding the expected topological signal of different model parameters.

In this paper, we show that summary statistics computed on the persistence diagram can be used for likelihood-based parametric inference.
We use genomic sequence data as a case study, examining the topological behavior of the coalescent process with recombination, a widely used stochastic model of biological evolution.
We find that the process generates nontrivial topology in a way that depends sensitively on parameter in the model.
The idea is presented as a proof of concept, in order to motivate the identification additional models with regular topological structure that may amenable to this type of inference.

\subsection{Related Work}

The application of persistent homology to genomic data was first introduced in \cite{Chan:2013vt}, where recombination rates in viral populations were estimated by computing $L_p$-norms on barcode diagrams.
The statistical properties of random simplicial complexes, including distributions over their Betti numbers, has been studied in \cite{Kahle:2011ep,Kahle:2013vy}.
The persistent homology of Gaussian random fields and other probabilistic structures has been studied in \cite{Adler:2010uy}.
Functions defined on the persistence diagram were used to compute a fractal dimension for various polymer physics models in \cite{MacPherson:2012eq}.

\section{Background}
\label{sec:background}

\subsection{Persistent Homology}

\begin{figure}
% \vskip 0.2in
\begin{center}
\centerline{\includegraphics[width=\columnwidth]{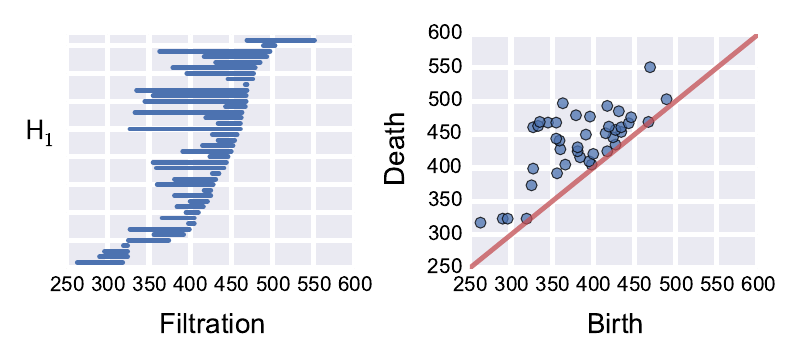}}
\caption{Two representations of the same topological invariants, computed using persistent homology. Left: Barcode diagram. Right: Persistence diagram. Data was generated from a coalescent simulation with $n=100$, $\rho=72$, and $\theta=500$.}
\label{fig:persistence_diagram}
\end{center}
\vskip -0.2in
\end{figure}

We summarize persistent homology from the perspective of an end-user.
For detailed background, see the reviews \cite{Carlsson:2009vh,Ghrist:2008tw} and the books \cite{Edelsbrunner:2010vl,Zomorodian:2005wf}.
In brief, persistent homology computes topological invariants representing information about the connectivity and holes in a dataset.
A dataset, $S=(s_{1},\ldots,s_{N})$, is represented as a point cloud in a high-dimensional space (not necessarily Euclidean).
From the point cloud, a nested family of simplicial complexes, or a filtration, is constructed, parameterized by a filtration value $\epsilon$, which controls the simplices present in the complex.
The two most common ways of constructing a simplicial complex at each $\epsilon$ are the \v{C}ech complex and the Vietoris-Rips complex.
The filtration is represented as a list of simplices defined on the vertices of $S$, annotated with the $\epsilon$ at which the simplex appears.
Given a filtration, the persistence algorithm is used to compute homology groups.
The $0$-dimensional homology ($H_0$) represents a hierarchical clustering of the data.
Higher dimensional homology groups represent loops, holes, and higher dimensional voids in the data.
Each feature is annotated with an interval, representing the $\epsilon$ at which the feature appears and the $\epsilon$ at which the feature contracts in the filtration.
These filtration values are the \emph{birth} and \emph{death} times, respectively.
The topological invariants in the filtration can be concisely represented in a barcode diagram, a set of line segments ordered by filtration value on the horizontal axis (Figure \ref{fig:persistence_diagram}).
Equivalently, invariants can represented by a persistence diagram, a scatter plot with the birth time on the horizontal axis and the death time on the vertical axis .
Persistent homology is computed using Dionysus \cite{Morozov:2012xx}.

\subsection{Coalescent Process}

The coalescent process is a stochastic model that generates the genealogy of individuals sampled from an evolving population \cite{Wakeley:2009ua}.
The genealogy is then used to simulate the genetic sequences of the sample.
This model is essential to many methods commonly used in population genetics.
Starting with a present-day sample of $n$ individuals, each individual's lineage is traced backward in time, towards a mutual common ancestor.
Two separate lineages collapse via a coalescence event, representing the sharing of an ancestor by the two lineages.
The stochastic process ends when all lineages of all sampled individuals collapse into a single common ancestor.
In this process, if the total (diploid) population size $N$ is sufficiently large, then the expected time before a coalescence event, in units of $2N$ generations, is approximately exponentially distributed:
\begin{equation}
P(T_{k}=t) \approx {k\choose2} e ^{-{k\choose2} t},
\end{equation}
where $T_k$ is the time that it takes for $k$ individual lineages to collapse into $k-1$ lineages.

After generating a genealogy, the genetic sequences of the sample can be simulated by placing mutations on the individual branches of the lineage.
The number of mutations on each branch is Poisson-distributed with mean $\theta t / 2$, where $t$ is the branch length and $\theta$ is the population-scaled mutation rate.
In this model, the average \emph{genetic distance} between any two sampled individuals, defined by the number of mutations separating them, is $\theta$.

The coalescent with recombination is an extension of this model that allows different genetic loci to have different genealogies.
Looking backward in time, recombination is modeled as a splitting event, occurring at a rate determined by population-scaled recombination rate $\rho$, such that an individual has a different ancestor at different loci.
Evolutionary histories are no longer represented by a tree, but rather by an \emph{ancestral recombination graph}.
Recombination is the component of the model generating nontrivial topology by introducing deviations from a contractibile tree structure, and is the component which we would like to quantify.
Coalescent simulations were performed using \texttt{ms} \cite{Hudson:2002vy}.

\begin{figure}
% \vskip 0.2in
\begin{center}
\centerline{\includegraphics[width=\columnwidth]{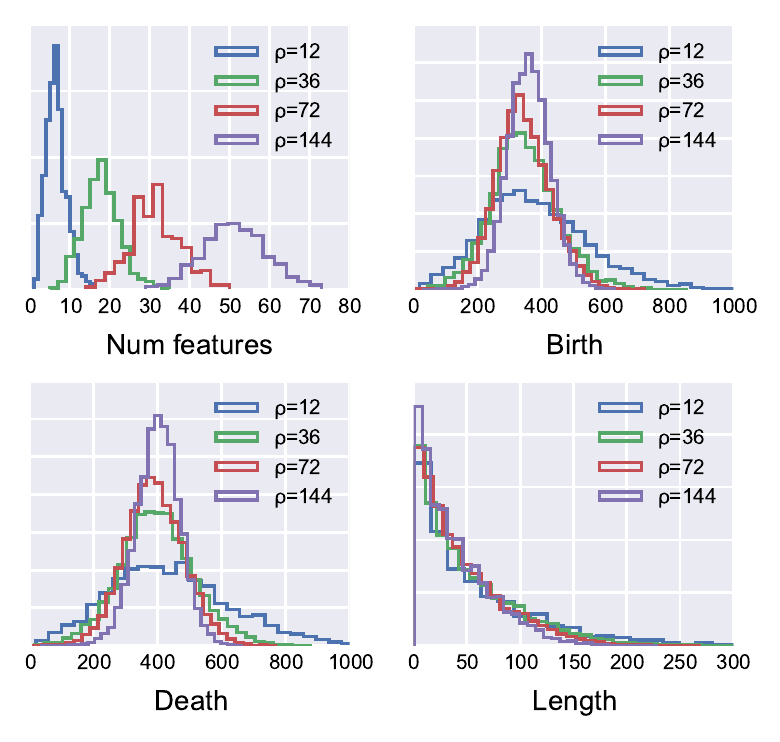}}
\caption{Distributions of statistics defined on the $H_1$ persistence diagram for different model parameters. Top left: Number of features. Top right: Birth time distribution. Bottom left: Death time distribution. Bottom right: Feature length distribution. Data generated from $1000$ coalescent simulations with $n=100$, $\theta=500$, and variable $\rho$.}
\label{fig:coalescent_sims}
\end{center}
\vskip -0.2in
\end{figure}

\section{Statistical Model}
\label{sec:model}

The persistence diagram from a typical coalescent simulation is shown in Figure \ref{fig:persistence_diagram}.
Examining the diagram, it would be difficult to classify the observed features into signal and noise.
Instead, we use the information in the diagram to construct a statistical model in order to infer the parameters, $\theta$ and $\rho$, which generated the data.
Note that we consider inference using only $H_1$ invariants, but the ideas easily generalize to higher dimensions.
We consider the following properties of the persistence diagram: the total number of features, $K$; the set of birth times, $(b_1,{\ldots},b_K)$; the set of death times, $(d_1,{\ldots},d_K)$; and the set of persistence lengths, $(l_1,{\ldots},l_K)$.
In Figure \ref{fig:coalescent_sims} we show the distributions of these properties for four values of $\rho$, keeping fixed $n=100$ and $\theta=500$.
Several observations are immediately apparent.
First, the topological signal is remarkably stable.
Second, higher $\rho$ increases the number of features, consistent with the intuition that recombination generates nontrivial topology in the model.
Third, the mean values of the birth and death time distributions are only weakly dependent on $\rho$ and are slightly smaller than $\theta$, suggesting that $\theta$ defines a natural scale in the topological space.
However, higher $\rho$ tightens the variance of the distributions.
Finally, persistence lengths are independent of $\rho$.

Examining Figure \ref{fig:coalescent_sims}, we can postulate: $K \sim \mathrm{Pois}(\zeta)$, $b_k \sim \mathrm{Gamma}(\alpha,\xi)$, and $l_k \sim \mathrm{Exp}(\eta)$.
Death time is given by $d_k=b_k+l_k$, which is incomplete Gamma distributed.
The parameters of each distribution are assumed to be an \emph{a priori} unknown function of the model parameters, $\theta$ and $\rho$, and the sample size, $n$.
Keeping $n$ fixed, and assuming each element in the diagram is independent, we can define the full likelihood as
\begin{equation}
p(D \given \theta,\rho) = p(K \given \theta,\rho)\displaystyle\prod_{k=1}^{K}p(b_k \given \theta,\rho)p(l_k  \given \theta,\rho).
\end{equation}
Simulations over a range of parameter values suggest the following functional forms for the parameters of each distribution.
The number of features is Poisson distributed with expected value
\begin{equation}
\zeta=a_{0}\log\left(1+\frac{\rho}{a_{1}+a_{2}\rho}\right)
\end{equation}
Birth times are Gamma distributed with shape parameter
\begin{equation}
\alpha=b_{0}\rho+b_{1}
\end{equation}
and scale parameter
\begin{equation}
\xi = \frac{1}{\alpha}(c_{0}\exp(-c_{1}\rho)+c_{2}).
\end{equation}
These expressions appears to hold well in the regime $\rho<\theta$, but break down for large $\rho$.
The length distribution is exponentially distributed with shape parameter proportional to mutation rate, $\eta=\alpha\theta$.
The coefficients in each of these functions are calibrated using simulations, and could be improved with further analysis.
This model has a simple structure and standard maximum likelihood approaches can be used to find optimal values of $\theta$ and $\rho$.

\section{Experiments}
\label{sec:experiments}

\subsection{Coalescent Simulations}

\begin{figure}[t]
% \vskip 0.2in
\begin{center}
\centerline{\includegraphics[width=\columnwidth]{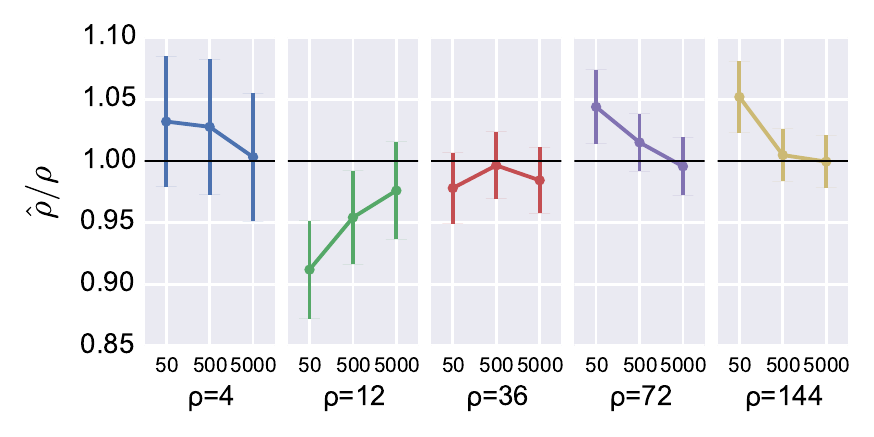}}
\caption{Inference of recombination rate $\rho$ using topological information. The recombination rate $\rho$ is estimated for five values \{4, 12, 36, 72, 144\} at three different mutation rates \{50, 500, 5000\}. Mean estimate over 500 simulations and 95\% confidence interval is shown.}
\label{fig:param_inference}
\end{center}
\vskip -0.2in
\end{figure}

We simulated a coalescent process with sample size $n=100$ and $l=10{,}000$ loci.
The mutation rate, $\theta$, was varied across $\theta=\{50,500,5000\}$.
The recombination rate, $\rho$, was varied across $\rho=\{4,12,36,72\}$.
The output of the process is a set of binary sequences of variable length (length is dependent on $\theta$).
The Hamming metric is used to construct a pairwise distance matrix between sequences.
We computed persistent homology and used the model described in Section \ref{sec:model} to estimate $\theta$ and $\rho$.
Results are shown in Figure \ref{fig:param_inference}, where we plot estimates and 95\% confidence interval from $500$ simulations.
We observe an improved $\rho$ estimate at higher mutation rate.
This is expected, as increasing $\theta$ is essentially increasing sampling on branches in the genealogy.
We also observe tighter confidence intervals at higher recombination rates, consistent with the behavior seen in Figure \ref{fig:coalescent_sims}.

% We compare our estimator of $\rho$ against a classical estimator \cite{Hudson:1987fo}.
% The classical estimator is based on the variance of pairwise distances in the sample, using observation that as $\rho$ increases, loci become uncorrelated, and the distance between samples begins to center around the mutation rate $\theta$.
% Coalescence and recombination are competing processes, and looking at the sequence divergence.
% In contrast, the observed $H_1$ invariants are strictly due to the presence of recombination.
% As shown in \cite{Chan:2013vt}, higher homology vanishes for models without recombination.

%(((Also include simulation with recombination at two scales?)))

\subsection{Application: Influenza Reassortment}

\begin{figure}
% \vskip 0.2in
\begin{center}
\centerline{\includegraphics[width=\columnwidth]{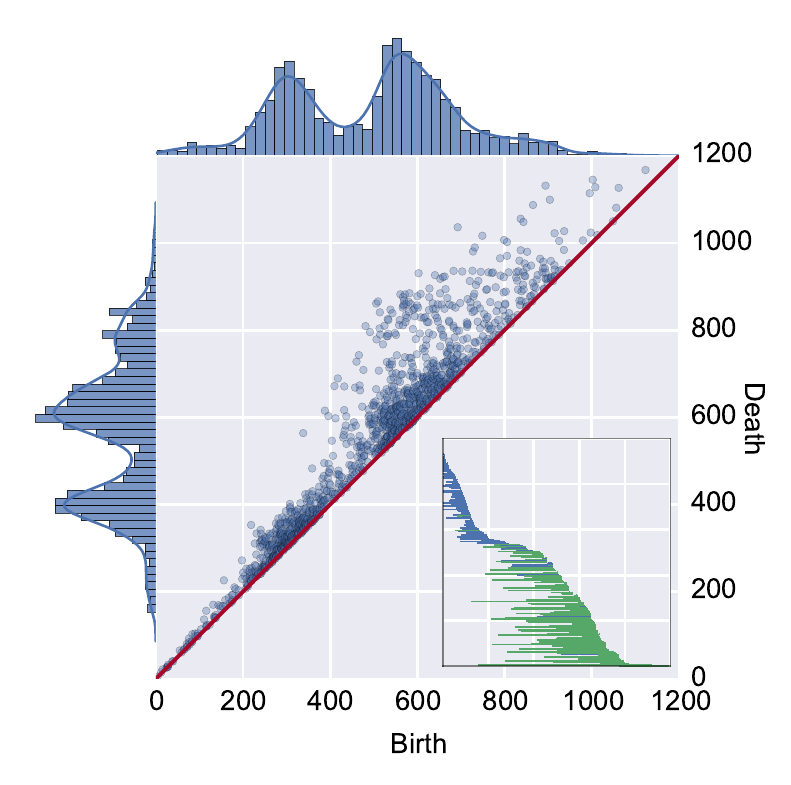}}
\label{fig:flu_scatterplot}
\caption{The $H_1$ persistence diagram computed from an avian influenza dataset. On the top and left are plotted the marginal distributions of birth and death times, along with a density estimate for each distribution. The bimodality indicates two scales of topological structure. Inset: The barcode diagram for a subset of this data. Blue bars have representative cycles involving only one subtype, green bars have cycles involving multiple subtypes.}
\end{center}
\vskip -0.2in
\end{figure}

To test our model on biological data, we considered reassortment in avian influenza virus.
Influenza is a single-stranded RNA virus that is naturally found in avian populations.
Each viral genome has eight genetic segments.
Subtypes are defined by two segments, hemagglutinin (HA) and neuraminidase (NA), e.g. H1N1 and H3N2.
When a host cell is coinfected with two different viral strains, reassortment of these segments can occur, such that viral offspring is a genetic mixture of the two parental strains.
Reassortment is of substantial medical interest, and has been connected with the outbreak of influenza epidemics.

We computed persistent homology on an aligned dataset of 3,105 avian influenza sequences across the seven major HA subtypes.
The persistence diagram is shown in Figure \ref{fig:flu_scatterplot}, along with density estimates for the birth and death distributions.
Both birth and death times appear strongly bimodal, unlike in the coalescent simulations, which were strictly unimodal.
This suggests two distinct scales of topological structure.
Using the representative cycles output by Dionysus on a subset of this data, we classified features as intrasubtype (involving one HA subtype) and intersubtype (involving multiple HA subtypes).
The $H_1$ barcode diagram for this data is shown in the Figure \ref{fig:flu_scatterplot} inset.
Intrasubtype features, in blue, occur at an earlier filtration scale than intersubtype features, in green.
The multiscale topological approach of persistent homology can distinguish biological events occuring at different genetic scales.

We isolated the two peaks and estimated two recombination rates: an intrasubtype $\rho_{1}=9.68$, and an intersubtype $\rho_{2}=21.43$.
We conclude that intersubtype recombination occurs at a rate over twice that of intrasubtype recombination, however a genetic barrier exists that maintains distinct subtype populations.
The nature of this barrier warrants further study.
This illustrates a real world example in which multiscale topological structure can be captured by persistent homology and given biological interpretation.

\section{Conclusions}
\label{sec:discussion}

In machine learning, the task is often to infer parameters of a model from observations.
This paper presented a proof of concept for statistical inference based on topological information computed using persistent homology.
Unlike previous work, which considered estimating homology of a partially observed object, we were interested in a model which generates a complex, but stable, topological signal.
Three conditions were required for the success of this approach:
First, a well-defined statistical model.
Second, an intuition that the observed topological structure is directly correlated with the parameters of interest in the model.
Third, sufficient topological signal to reliably estimate statistics on the persistence diagram.
It is an open question to identify classes of models for which these conditions will hold.

\bibliography{icml_manuscript}
\bibliographystyle{icml2014}

\end{document}